\documentclass[hyper]{prop2015}
\usepackage[english]{babel}

\usepackage{graphicx}
\usepackage{psfrag}
\usepackage{amsmath,bbm}
\usepackage{amsthm}
\usepackage{amssymb}
\usepackage{epsfig}
\usepackage{euscript}
\usepackage{array}
\usepackage{cancel}
\usepackage{mathtools}
\usepackage{empheq}
\usepackage{graphicx}
\usepackage{subfigure}
\usepackage{upgreek}
\usepackage{epstopdf}
\usepackage{booktabs}
\usepackage{mathrsfs}
\usepackage{amsbsy}
\usepackage{hyperref}

\newcommand{\be}{\begin{equation}}
\newcommand{\ee}{\end{equation}}
\newcommand{\eq}[1]{(\ref{#1})}
\newcommand{\bit}{\begin{itemize}}  \newcommand{\eit}{\end{itemize}}
\newcommand{\ben}{\begin{enumerate}}  \newcommand{\een}{\end{enumerate}}

\newcommand{\bm}[1]{\mbox{\boldmath $#1$}}
\newcommand{\rf}[1]{(\ref{#1})}

\def\bd{\begin{document}}
\def\ed{\end{document}}
 \def\bea{\begin{eqnarray}}
 \def\eea{\end{eqnarray}}
\let\bm=\bibitem

\def\la{\langle}
\def\ra{\rangle}

\def\npb#1#2#3{Nucl. Phys. {\bf{B#1}} #3 (#2)}
\def\plb#1#2#3{Phys. Lett. {\bf{#1B}} #3 (#2)}
\def\prl#1#2#3{Phys. Rev. Lett. {\bf{#1}} #3 (#2)}
\def\prd#1#2#3{Phys. Rev. {D bf{#1}} #3 (#2)}
\def\cmp#1#2#3{Comm. Math. Phys. {\bf{#1}} #3 (#2)}
\def\cqg#1#2#3{Class. Quantum Grav. {\bf{#1}} #3 (#2)}
\def\nppsa#1#2#3{Nucl. Phys. B (Proc. Suppl.) {\bf{#1A}}#3 (#2)}
\def\ap#1#2#3{Ann. of Phys. {\bf{#1}} #3 (#2)}
\def\ijmp#1#2#3{Int. J. Mod. Phys. {\bf{A#1}} #3 (#2)}
\def\rmp#1#2#3{Rev. Mod. Phys. {\bf{#1}} #3 (#2)}
\def\mpla#1#2#3{Mod. Phys. Lett. {\bf A#1} #3 (#2)}
\def\jhep#1#2#3{J. High Energy Phys. {\bf #1} #3 (#2)}
\def\atmp#1#2#3{Adv. Theor. Math. Phys. {\bf #1} #3 (#2)}

\def\N{{\cal N}}
\def\sst{\scriptscriptstyle}
\def\thetabar{\bar\theta}
\def\Tr{{\rm Tr}}
\def\one{\mbox{1 \kern-.59em {\rm l}}}

\def\a{\alpha}      \def\da{{\dot\alpha}}  \def\dA{{\dot A}}
\def\b{\beta}       \def\db{{\dot\beta}}
\def\g{\gamma}  \def\G{\Gamma}  \def\dc{{\dot\gamma}}
\def\d{\delta}  \def\D{\Delta}  \def\ddt{\dot\delta}
\def\e{\epsilon}
\def\ve{\varepsilon}
\def\uve{\upvarepsilon}
\def\f{\phi}    \def\F{\Phi}    \def\vvf{\f}
\def\vphi{\varphi}
\def\h{\eta}
\def\k{\kappa}
\def\l{\lambda} \def\L{\Lambda}
\def\m{\mu} \def\n{\nu}
\def\o{\omega}
\def\p{\pi} \def\P{\Pi}
\def\r{\rho}
\def\s{\sigma}  \def\S{\Sigma}
\def\t{\tau}
\def\th{\theta} \def\Th{\Theta} \def\vth{\vartheta}
\def\X{\Xeta}
\def\z{\zeta}

\def\na{\nabla}

\def\cA{{\mathcal A}} \def\cB{{\cal B}} \def\cC{{\cal C}}
\def\cD{{\cal D}} \def\cE{{\cal E}} \def\cF{{\cal F}}
\def\cG{{\cal G}} \def\cH{{\cal H}} \def\cI{{\cal I}}
\def\cJ{{\mathscr J}} \def\cK{{\cal K}} \def\cL{{\cal L}}
\def\cM{{\cal M}} \def\cN{{\cal N}} \def\cO{{\cal O}}
\def\cP{{\cal P}} \def\cQ{{\cal Q}} \def\cR{{\cal R}}
\def\cS{{\cal S}} \def\cT{{\cal T}} \def\cU{{\cal U}}
\def\cV{{\cal V}} \def\cW{{\cal W}} \def\cX{{\cal X}}
\def\cY{{\cal Y}} \def\cZ{{\cal Z}}

\def\ua{\underline{\alpha}}
\def\uc{\underline{\phantom{\alpha}}\!\!\!\gamma}
\def\um{\underline{\mu}}
\def\ud{\underline\delta}
\def\ue{\underline\epsilon}
\def\una{\underline a}\def\unA{\underline A}
\def\unb{\underline b}\def\unB{\underline B}
\def\unc{\underline c}\def\unC{\underline C}
\def\und{\underline d}\def\unD{\underline D}
\def\une{\underline e}\def\unE{\underline E}
\def\unf{\underline{\phantom{e}}\!\!\!\! f}\def\unF{\underline F}
\def\unm{\underline m}\def\unM{{\underline M}}
\def\unn{\underline n}\def\unN{{\underline N}}
\def\unp{\underline{\phantom{a}}\!\!\! p}\def\unP{\underline P}
\def\unq{\underline{\phantom{a}}\!\!\! q}
\def\unQ{\underline{\phantom{A}}\!\!\!\! Q}
\def\unH{\underline{H}}

\def\As {{A \hspace{-6.4pt} \slash}\;}
\def\bs {{b \hspace{-6.4pt} \slash}\;}
\def\Ds {{D \hspace{-6.4pt} \slash}\;}
\def\Gts {{\Gt \hspace{-6.4pt} \slash}\;}
\def\ds {{\del \hspace{-6.4pt} \slash}\;}
\def\ss {{\s \hspace{-6.4pt} \slash}\;}
\def\ks {{ k \hspace{-6.4pt} \slash}\;}
\def\ps {{p \hspace{-6.4pt} \slash}\;}
\def\xs {{x \hspace{-6.4pt} \slash}\;}
\def\pas {{{p_1} \hspace{-6.4pt} \slash}\;}
\def\pbs {{{p_2} \hspace{-6.4pt} \slash}\;}
\def\cFs {{{\cal F} \hspace{-6.4pt} \slash}\;}
\def\Dss {{D \hspace{-7.5pt} \slash}\;}
\def\dss {{\del \hspace{-7.0pt} \slash}\;}

\def\Ah{{\hat{A}}}
\def\Ch{{\hat{C}}}
\def\Dh{{\hat{D}}}
\def\Gh{{\hat{G}}}
\def\Fh{{\hat{F}}}
\def\Ih{{\hat{I}}}
\def\Jh{{\hat{J}}}
\def\Kh{{\hat{K}}}
\def\Lh{{\hat{L}}}
\def\Ph{{\hat{P}}}
\def\Rh{{\hat{R}}}
\def\Vh{{\hat{V}}}
\def\Xh{{\hat{X}}}

\def\ah{{\hat{\a}}}
\def\bh{{\hat{\b}}}
\def\gh{{\hat{\g}}}
\def\dh{{\hat{\d}}}
\def\rh{{\hat{\r}}}
\def\hh{\hat{h}}
\def\uh{\hat{u}}
\def\xh{\hat{x}}
\def\yh{\hat{y}}
\def\ph{\hat{p}}
\def\xih{\hat{\xi}}
\def\chih{\hat{\chi}}
\def\Psih{\hat{\Psi}}
\def\phih{\hat{\phi}}

\def\psit{\tilde{\psi}}
\def\Psit{\tilde{\Psi}}
\def\Psibt{\tilde{\bar{Psi}}}

\def\st{\tilde{\sigma}}

\def\delt{\tilde{\delta}}
\def\Phit{\tilde{\Phi}}
\def\Phitb{\overline{\tilde{Phi}}}
\def\tht{\tilde{\th}}
\def\lt{\tilde{\l}}
\def\chit{\tilde{\chi}}
\def\phit{\tilde{\phi}}

\def\At{\tilde{A}}
\def\Bt{\tilde{B}}
\def\Ct{\tilde{C}}
\def\Dt{\tilde{D}}
\def\Et{\tilde{E}}
\def\Ft{\tilde{F}}
\def\Gt{\tilde{G}}
\def\Ht{\tilde{H}}
\def\It{\tilde{I}}
\def\Jt{\tilde{J}}
\def\Qt{\tilde{Q}}
\def\Rt{\tilde{R}}
\def\Mt{\tilde{M }}
\def\Nt{\tilde{N}}
\def\St{\tilde{S}}
\def\Vt{\tilde{V}}
\def\Xt{\tilde{X}}
\def\at{\tilde{a}}
\def\ct{\tilde{c}}
\def\dt{\tilde{d}}
\def\htt{\tilde{h}}
\def\ft{\tilde{f}}
\def\gt{\tilde{g}}
\def\pt{\tilde{p}}
\def\qt{\tilde{q}}
\def\vt{\tilde{v}}
\def\nt{\tilde{n}}
\def\ut{\tilde{u}}
\def\wt{\tilde{w}}
\def\zt{\tilde{z}}
\def\xt{\tilde{x}}
\def\yt{\tilde{y}}
\def\Psit{\tilde{\Psi}}
\def\vphit{\tilde{\varphi}}
\def\tD{\tilde{\D}}

\def\eb{\bar{\epsilon}}
\def\delb{\bar{\partial}}
\def\thb{\bar{\theta}}
\def\mub{\bar{\mu}}
\def\lamb{\bar{\l}}
\def\psib{\bar{\psi}}
\def\sb{\bar{\sigma}}
\def\xib{\bar{\xi}}
\def\chib{\bar{\chi}}

\def\Psib{\bar{\Psi}}
\def\Phib{\bar{\Phi}}
\def\Lamb{\bar{\Lambda}}
\def\Sb{{\overline \Sigma}}
\def\cb{\bar{c}}
\def\hb{\bar{h}}
\def\qb{\bar{q}}
\def\wb{\bar{w}}
\def\ub{\bar{u}}
\def\zb{{\bar{z}}}
\def\Hb{\bar{H}}
\def\Qb{{\bar Q}}
\def\Omegab{\overline{\Omega}}
\def\ob{\overline{\omega}}

\def\Ab{{\overline A}} \def\Bb{{\overline B}} \def\Cb{{\overline C}}
\def\Db{{\overline D}} \def\Eb{{\overline E}} \def\Fb{{\overline F}}
\def\Gb{{\overline G}}
\def\Ib{{\overline I}}
\def\Jb{{\overline J}} \def\Kb{{\overline K}} \def\Lb{{\overline L}}
\def\Mb{{\overline M}} \def\Nb{{\overline N}} \def\Ob{{\overline O}}
\def\Pb{{\overline P}}  \def\Rb{{\overline R}}
 \def\Tb{{\overline T}} \def\Ub{{\overline U}}
\def\Vb{{\overline V}} \def\Wb{{\overline W}} \def\Xb{{\overline X}}
\def\Yb{{\overline Y}} \def\Zb{{\overline Z}}

\def\fb{{\overline f}}
\def\gb{{\overline g}}
\def\mb{{\overline m}}
\def\lb{{\overline l}}
\def\yb{{\overline y}}

\def\ba{{\bf a}}
\def\bk{{\bf k}}
\def\bl{{\bf l}}
\def\bp{{\bf p}}
\def\bq{{\bf q}}
\def\br{{\bf r}}
\def\bt{{\bf t}}
\def\bu{{\bf u}}
\def\bv{{\bf v}}
\def\bx{{\bf x}}
\def\by{{\bf y}}
\def\bA{{\bf A}}
\def\bB{{\bf B}}
\def\bR{{\bf R}}
\def\bV{{\bf V}}

\def\bz{{\boldsymbol{\zeta}}}

\def\bone{{\bf 1}}

\def\va{{\vec a}}
\def\vk{{\vec k}}
\def\vp{{\vec p}}
\def\vq{{\vec q}}
\def\vx{{\vec x}}
\def\vy{{\vec y}}
\def\vu{{\vec u}}
\def\vv{{\vec v}}
\def \vH{{\vec H}}
\def \vg{{\vec g}}

\def\vs{{\vec \sigma}}
\def\vtau{{\vec \tau}}

\def\frA{\mathfrak{A}}
\def\frB{\mathfrak{B}}
\def\frC{\mathfrak{C}}
\def\frD{\mathfrak{D}}
\def\frE{\mathfrak{E}}
\def\frF{\mathfrak{F}}
\def\frG{\mathfrak{G}}
\def\frH{\mathfrak{H}}
\def\frM{\mathfrak{M}}
\def\frN{\mathfrak{N}}
\def\frR{\mathfrak{R}}
\def\frW{\mathfrak{W}}

\def\fra{\mathfrak{a}}
\def\frb{\mathfrak{b}}
\def\frf{\mathfrak{f}}
\def\frg{\mathfrak{g}}
\def\frh{\mathfrak{h}}
\def\frl{\mathfrak{l}}
\def\frs{\mathfrak{s}}
\def\fri{\mathfrak{i}}
\def\frj{\mathfrak{j}}

\def\ma{\mathfrak{a}}
\def\mg{\mathfrak{g}}
\def\mh{\mathfrak{h}}
\def\mR{\mathfrak{R}}
\def\mN{\mathfrak{N}}

\newcommand{\nn}{{\nonumber}}

\def\d{\delta}\def\D{\Delta}\def\ddt{\dot\delta}

\def\pa{\partial} \def\del{\partial}
\def\xx{\times}
\def\uno{\mbox{1 \kern-.59em {\rm l}}}

\def\trp{^{\top}}
\def\inv{^{-1}}
\def\dag{\dagger}
\def\pr{^{\prime}}

\def\rar{\rightarrow}
\def\lar{\leftarrow}
\def\lrar{\leftrightarrow}

\newcommand{\0}{\,\!}      
\def\one{1\!\!1\,\,}
\def\im{\imath}
\def\jm{\jmath}

\newcommand{\tr}{\mbox{tr}}
\newcommand{\slsh}[1]{/ \!\!\!\! #1}

\def\vac{|0\rangle}
\def\lvac{\langle 0|}

\def\hlf{\frac{1}{2}}
\def\ove#1{\frac{1}{#1}}
\newcommand{\hot}[1]{\frac{#1}{2}}

\def\Box{\square}
\def\CC {\mathbb{C}}
\def\FF {\mathbb{F}}
\def\RR{\mathbb{R}}
\def\NN{\mathbb{N}}
\def\ZZ{\mathbb{Z}}
\def\bb#1{{\bf #1}}
\def\bcomment#1{}
\def\bfhat#1{{\bf \hat{#1}}}
\def\VEV#1{\left\langle #1\right\rangle}

\newcommand{\ex}[1]{{\rm e}^{#1}} \def\ii{{\rm i}}

\newcommand{\lrbrk}[1]{\left(#1\right)}
\newcommand{\lrsbrk}[1]{\left[#1\right]}
\newcommand{\sfrac}[2]{{\textstyle\frac{#1}{#2}}}

\def\stw{{\sqrt{2}}}

\def\rf {{\rm f}}
\def\ri {{\rm i}}
\def\rj {{\rm j}}
\def\rn {{\rm n}}
\def\rk {{\rm k}}
\def\rl {{\rm l}}
\def\rr {{\rm r}}
\def\rs {{\scriptscriptstyle \rm S}}
\def\rt {{\scriptscriptstyle \rm T}}

\def\rQ {{\scriptscriptstyle \rm \cQ}}
\def\rR {{\scriptscriptstyle \rm \cR}}

\def\cQb{{\cal \Qb}}
\def\cRb{{\cal \Rb}}
\def\cWb{{\cal \Wb}}

\def\fd {{\rm N}}
\def\afd {{\overline{\rm N}}}

\def \II {I\hspace{-.1em}I\hspace{.1em}}
\def \IIA {\mbox{\II A\hspace{.2em}}}
\def \IIB {\mbox{\II B\hspace{.2em}}}
\def \gs {g^s}
\def \ls {\lambda^s}

\def \I {{\cal I}}
\def \qs {q\hspace{-.53em}/\hspace{.15em}}
\def \ks {k\hspace{-.53em}/\hspace{.15em}}
\def \YM {{\mbox{\tiny YM}}}
\def \gym {g_{\YM}}

\def \Lc {\L_c}
\def\IR{\relax{\rm I\kern-.18em R}}
\def \id {{\bf 1}}

\def\cci{\ell}
\def\ccj{\ell'}

\def\bbq{\pmb{q}}

\newcommand{\CN}{\mathcal{N}}
\category{Proceedings}
\keywords{M-branes, string theory, boundary conformal field theory}
\subtitle{\href{http://www.maths.dur.ac.uk/lms/109/index.html}{LMS/EPSRC Durham Symposium on Higher Structures in M theory}}
\title{Weyl Anomaly and Vacuum Magnetization Current of M5-brane in Background Flux}

\author[C.-S. Chu]{Chong-Sun Chu\inst{a,b,}\footnote{Corresponding author e-mail:~\href{mailto:cschu@phys.nthu.edu.tw}{\textsf{cschu@phys.nthu.edu.tw}}}}
\address[1]{Physics Division, National Center for Theoretical
  Sciences, \\
National Tsing-Hua University, Hsinchu, 30013, Taiwan}
\address[2]{Department of Physics, National Tsing-Hua University, Hsinchu 30013, Taiwan}
\shortabstract
\begin{abstract}
It was recently discovered that for a boundary system
  in the presence of a background
  magnetic field, the quantum fluctuation of the vacuum would
  create a non-uniform magnetization density for the vacuum and a magnetization
  current is induced in the vacuum.
  It was also shown that this `magnetic
  Casimir effect' of the vacuum is closely related to another quantum
  effect of the
  vacuum,  the Weyl anomaly. Furthermore, the phenomena can be
  understood in terms of the holography of the
  boundary system .
In this article, we review the derivation of this phenomena from QFT 
as well as the derivation of it using AdS/BCFT.
 We then generalize this four dimensional
  effect to six-dimensions. We use the AdS/BCFT holography to show  
  that  in the presence of a 3-form magnetic
  field strength $H$, 
  a string current is induced in a six-dimensional boundary conformal field
  theory. This  allows us to determine the gauge field contribution to the
  Weyl anomaly in six-dimensional  conformal field theory in a
  $H$-flux background.
  For the (2,0) superconformal field theory of $N$ M5-branes, the
  current has a magnitude proportional to $N^3$ for large $N$.
  This suggests that
  the degree of freedoms scales as $N^3$ in the (2,0) superconformal
  theory of $N$ multiple M5-branes. Our result for the Weyl anomaly is a new prediction for the (2,0) theory.
  \end{abstract}
\begin{document}
\maketitle


\section{M5-branes}

The decoupling limit of $N$ coincident M5-branes 
is given by an interacting (2,0) superconformal
theory in six-dimensions \cite{Witten:1995zh}.
The understanding of the dynamics
of this system is of utmost importance. It will not only improve our understanding
of the AdS/CFT correspondence for the AdS$_7\times S^4$ background \cite{Maldacena:1997re}; 
in addition, as the problem involves a mathematical 
formulation of a self-duality equation
for a non-Abelian 3-form gauge field strength, one may suspect that it may have
an impact on mathematical physics in  a way similar to its 
lower-dimensional cousin, the self-dual Yang--Mills equation.\footnote{Some of the important mathematical results and applications are, for example, the ADHM construction of instantons \cite{Atiyah:1978ri}, the Hitchin integrable system \cite{Hitchin:1987:91-114} and the classification of 4-manifolds \cite{0198502699}.}

On general grounds, the theory of multiple M5-branes does not have
a free dimensionless parameter and is inherently non-perturbative. 
It does not mean
that an action does not exist, though it does mean that the action will
be of limited use, probably no more than giving the 
corresponding equation of motion.
This is still very interesting since one can expect that  
non-trivial space-time physics of M-theory could be learned from the physics of the 
solitonic objects of the world-volume theory of M5-branes, much like 
the cases of M2-branes and D-branes. See for example, \cite{Tong:2005un}.

In the paper \cite{Chu:2012um}, a consistent self-duality equation of 
motion for a non-Abelian tensor gauge field in six-dimensions has been 
constructed and proposed to be the low energy equation of motion of the
self-dual tensor field living on the world-volume of a system of 
multiple M5-branes. 
The self-dual equation of motion proposed in \cite{Chu:2012um} is meant to be an 
effective description for the M5-branes in the long length limit,
just like the supergravity equation of motion provides
an effective description for the M-theory. 
The non-Abelian self-duality equation constructed in \cite{Chu:2012um} 
generalizes the equation of motion
for a single M5-brane of \cite{Howe:1996yn,Howe:1997fb,Perry:1996mk,Aganagic:9701166,Bandos:9703127,Bandos:9701149}. 
It  was constructed 
in the gauge $B_{5\m} =0$ ($\m =0,\cdots, 4$) and is a non-Abelian 
generalization of the 
Henneaux--Teitelboim--Perry--Schwarz construction for the $U(1)$ case \cite{Perry:1996mk,Henneaux:1988gg}.
The construction of \cite{Chu:2012um} involves the introduction of a set of 
non-propagating non-Abelian 1-form gauge fields which was motivated originally 
by the boundary analysis  in \cite{Chu:2009ms} and  further analyzed
 \cite{Chu:2011fd}. This aspect is very similar to the BLG \cite{Bagger:2006sk,Gustavsson:2007vu,Bagger:2007jr}
and ABJM model \cite{Aharony:2008ug} of multiple M2-branes where a set of
non-propagating Chern--Simons gauge fields was introduced in order to
allow for a simple representation of
the highly non-linear and non-local self interactions of the matter fields
of the theory.

The proposed self-duality equation  reads
\begin{subequations}
\be\label{sd-na}
\Ht_{\m\n} = \pa_5 B_{\m\n},
\ee
where the gauge field $A_\m$ is constrained to be  given by
\be \label{FH}
F_{\m\n} = c\int {\rm d}x_5\, \Ht_{\m\n}.
\ee
Here
\be
H_{\m\n\r}= D_{[\m}B_{\n\r]} = \pa_{[\m}B_{\n\r]}+[A_{[\m},B_{\n\r]}],
\ee
\be
\tilde{H}_{\m\n} = \ove{6}\e_{\m\n\r\s\t}H^{\r\s\t},\qquad \e_{01234}=-1,
\ee
\be
F_{\m\n} = \pa_{\m} A_{\n} - \pa_{\n} A_{\m} + [A_\m,A_\n].
\ee
\end{subequations}
All fields are in the adjoint representation of the Lie algebra of 
the gauge group $G$, and $c$ is a free parameter. 

Evidence that this self-duality equation describes
the physics of multiple M5-branes was provided in \cite{Chu:2012um}, and further
in \cite{Chu:2012rk,Chu:2013hja,Chu:2013joa,Chu:2013gra}.
In
\cite{Chu:2012rk,Chu:2013hja}, non-Abelian self-dual string solutions were constructed
and a precise agreement \cite{Chu:2013hja} of the 
field theory results and the supergravity descriptions \cite{Niarchos:2012pn,Niarchos:2013ia} 
was found. Moreover it was found that the constant $c$ is 
fixed by quantization condition of the self-dual strings solution 
of the theory: 
\be
c = \sqrt{N (N+1)} (N-1) \sim N^2,
\ee
where the second relation holds for large $N$. 
This is satisfying  as otherwise $c$ would be 
a free dimensionless constant in the theory and hence contradicts with
what we know about M5-branes in flat space.
One thing interesting about the self-dual string 
solutions constructed in \cite{Chu:2012rk,Chu:2013hja} is that the auxiliary gauge field
is always given by a magnetic monopole  which gives rise
to the charge of the self-dual string.
This was shown to be case for the original 
Perry--Schwarz self-dual string and the Wu--Yang self-dual string \cite{Chu:2012rk},
as well as for the generalized Wu--Yang self-dual string \cite{Chu:2013hja}, with the
corresponding monopole configurations given by the Dirac monopole, 
the Wu--Yang monopole and the generalized Wu--Yang monopole. 
In the paper \cite{Chu:2013gra}, the construction of self-dual string solutions to the non-Abelian two-form self-duality equation proposed in \cite{Chu:2012um} was generalized to cover the general case of having a   four dimensional non-Abelian BPS monopoles in its core.
The self dual string charge is given by the charge of the monopole. The construction suggests a Nahm-like
construction for non-Abelian self-dual string, which has been 
speculated and analyzed by other authors \cite{Saemann:2010cp,Palmer:2011vx,Campos:2000de}. 
In \cite{Chu:2013joa}, non-Abelian wave configurations 
which are supported by Yang--Mills instanton were constructed
and they were found to match up precisely with the description of 
M-wave on the world-volume of M5-branes system.

There exists a number of  proposals for the fundamental formulation of the
six-dimensional (2,0)
theory: most notably, these include the discrete light-cone quantisation
definition
based on quantum mechanics on the moduli space of instantons \cite{Aharony:1997th,Aharony:1997an},
a definition based on deconstruction from four dimensional superconformal,
quiver field theories \cite{ArkaniHamed:2001ie}, and the conjecture that the (2,0) theory
compactified on a circle is equivalent to the five-dimensional maximally
supersymmetric Yang--Mills theory \cite{Douglas:2010iu,Lambert:2010iw}. And despite
an extensive amount
of work on this topic, see for example, \cite{Chu:2011fd,Ho:2011ni,
Samtleben:2011fj,Kim:2011mv,Kallen:2012zn,Bern:2012di,
Cordova:2015vwa,Beem:2015aoa,Hosomichi:2012ek,Kallen:2012va,Kim:2012ava,Kim:2012gu,Kim:2012qf,Saemann:2012uq,Saemann:2016sis},
the multiple M5-branes
theory remains mysterious. 
In addition to consistency and symmetry
requirement, the fundamental theory, no matter how
it is defined, should reproduce properties that are expected of
the multiple M5-branes system. For example, it should describe a
non-trivial interacting theory of (2,0) superconformal multiplets. It should
contain BPS states of self-dual strings which corresponds to boundaries
of M2-branes ending on the stack of M5-branes
\cite{Strominger:1995ac,Dijkgraaf:1996hk}.
It should explain the S-duality of the
$\cN=4$ supersymmetric Yang--Mills theory \cite{Tachikawa:2011ch}.
It should also make apparent the $N^3$
entropy behaviour \cite{Klebanov:1996un}. In particular it should explain
whether this is due to novel degrees of freedom of the (2,0) theory or not.

To this end, a novel approach was adopted in the paper \cite{Chu:2018fpx}. There a boundary was introduced 
to the M5-brane system. By using a holographic duality for BCFT \cite{Takayanagi:2011zk,Miao:2017gyt,Chu:2017aab}, it was found that   
in the presence of a 3-form magnetic  field strength $H$, 
  a string current is induced in a six-dimensional boundary conformal field
  theory. This  allows us to determine the gauge field contribution to the
  Weyl anomaly in six-dimensional  conformal field theory in a
  $H$-flux background.  For the (2,0) superconformal field theory of $N$ M5-branes,
the current has a magnitude proportional to $N^3$ for large $N$.
This suggests that the
degree of freedoms scales as $N^3$ in the (2,0) superconformal theory of $N$ multiple
M5-branes
\footnote{
An $N^3$ dependence has also been found for the conformal $a$-anomaly, the Euler density
contribution, by relating the six-dimensional 
 the Coulomb 
branch of the (2, 0) theory with the Coulomb branch interactions in four dimensions using supersymmetry \cite{Ganor:1997jx,Maxfield:2012aw}.}. 
 The prediction we have for the induced string current and the Weyl anomaly is a new criteria that the (2,0) theory should satisfy.

\section{Holographic boundary current}

\subsection{Induced particle current}
It was recently discovered that for a boundary system
  in the presence of a background
  magnetic field, the quantum fluctuation of the vacuum would
  create a non-uniform magnetization density for the vacuum and a magnetization
  current is induced in the vacuum \cite{Chu:2018ksb}.
  It was also shown that this `magnetic
  Casimir effect' of the vacuum is closely related to another quantum
  effect of the
  vacuum,  the Weyl anomaly. Furthermore, the phenomena can be
  understood in terms of the holography of the
  boundary system \cite{Chu:2018ntx}. We review the derivations of this result 
 below. 
  
   In general, for a boundary quantum field theory (BQFT),
the renormalized current is generally singular near
the boundary and the expectation value takes the asymptotic form:
\begin{eqnarray}\label{current0}
\la J_\m\ra = x^{-3} J^{(3)}_{\m}+x^{-2} J^{(2)}_{\m}+x^{-1} J^{(1)}_{\m},
\quad x \sim 0,
\end{eqnarray}
where $x$ is the proper distance from the boundary and $J^{(n)}_{\m}$
depend only the background geometry, the background vector field strength
and the type of fields under consideration.
Hereafter we will drop the symbol $\la \; \ra$ for the expectation value. 
A 
similar expansion has been considered for the renormalized stress tensor
\cite{Deutsch:1978sc}.
We consider current that is conserved
\begin{eqnarray}\label{divergenceless}
  D_\m  J^\m  = 0
\end{eqnarray}
up to  possibly an anomaly term. Since this
term
is finite,
it is irrelevant to the divergent part of renormalized current
(\ref{current0}).  Substituting (\ref{current0}) into
(\ref{divergenceless}), we obtain
the gauge invariant solution
\begin{equation}\label{solnJ1}
\begin{split}
  &J^{(3)}_{ \m}=0, \quad \ J^{(2)}_{\m}=0,\\ &J^{(1)}_{\m}=
  \a_1  F_{\m\n} n^\n
  +\a_2 \mathcal{D}_\m k+\a_3 \mathcal{D}_\n k^\n_\m
 +\a_4  \star F_{\m\n}\, n^\n
\end{split}
\end{equation}
where
$F_{\m\n}$, $\star F_{\m\n}$, $n_\m$, $\mathcal{D}_m$, $k_{\m\n}$ and
$h_{\m\n}$ are respectively the background field strength, 
Hodge dual of field strength, 
the normal vector, induced
covariant derivative, extrinsic curvature and induced metric of the
boundary. 
Note that  in \eq{solnJ1} we have re-expressed
$n^\m R_{\m\n}h^{\m}_\n$ in terms of extrinsic
curvatures by using the Gauss--Codazzi equation
$n^\m R_{\m\n}h^{\n}_\g =\mathcal{D}_\m k^\m_\g- \mathcal{D}_\g k$.
Here the coefficients $\a_i$ are arbitrary and
the expression \eq{solnJ1} gives the most
general form of boundary behavior of the
current that is consistent with the conservation law and gauge invariance.

 Consider a conformal field theory (CFT) with partition function $Z[g_{\m\n}]$ and the effective action
$W[g_{\m\n}] = \ln Z[g_{\m\n}]$. 
The scaling symmetry of CFT is generally broken due to quantum effects
and the breaking is measured by the Weyl anomaly
\be
\mathcal{A} := \del_\vphi W[e^{2 \vphi} g_{\m\n}] \big|_{\vphi=0} = \int_M \la
T_\m^\m \ra.
\ee
The metric contribution to the Weyl anomaly is well understood. For example in
even dimensions, the bulk part of the Weyl anomaly takes the form
\cite{Deser:1993yx}
\be
\la T_\m^\m \ra = \frac{1}{(4\pi)^{d/2}} \left(
\sum_j c_{dj}I^{(d)}_j - (-1)^{\frac{d}{2}}a_d E_d
\right).
\ee
Here 
$E_d$ is the Euler density in $d$ dimensions,
$I^{(d)}_j$ are independent Weyl invariants of weight $-d$ and the subscript $j$
labels the Weyl invariants. The boundary terms of the Weyl anomaly has also been
studied and classified recently in \cite{Herzog:2015ioa}. In general, in addition
to a nontrivial background metric, one may also turn on a gauge field background
and the loops of matter fields  will
give  a Weyl anomaly. For example in four dimensions,
vector gauge field (Abelian or non-Abelian)
is classically conformal and there is a  Weyl anomaly \cite{Peskin:1995ev}
\be \label{anom-4d}
\la T_\m^\m \ra = b\; \tr F^2.
\ee
Here $b = \beta(g)/2g^3$ and $\beta(g)$ is the beta function of the theory
$S= -1/(4g^2) \int \tr F^2$.

In \cite{Chu:2018ntx} the following
relation was observed and proven:
\be \label{key}
(\d \cA )_{\del M} = \Big(\int_M \sqrt{g} J^\mu \d A_\mu
\Big)_{\log \frac{1}{\e}},
\ee
where a regulator $x \geq \e$ to the boundary is introduced for the integral
on the
right hand side of \eq{key}. The relation \eq{key} holds in general for a boundary conformal field theory (BCFT)
\cite{Miao:2017aba} and
identifies the boundary contribution of
the variation of the Weyl anomaly
under an arbitrary variation of the gauge field $\d A_\m$ with  the UV
logarithmic divergent part of the integral
involving the expectation value
$J^\m$ of the renormalized $U(1)$ current.

Using the relation \eq{key} one can fix the current coefficients in
terms of the boundary central charges of the theory.
For four-dimensional unitary quantum field theories (QFTs) without the parity
odd anomaly term, one has
\begin{eqnarray}
  \label{main1}
\a_1=4b_1, \quad \a_2 = \a_3 = \a_4 = 0
\end{eqnarray}
and hence the vacuum expectation value of the current
\begin{eqnarray}
  \label{current2}
 J_b  = \frac{4 b_1 F_{b n}}{x}, \quad x \sim 0,
\end{eqnarray}
near the boundary. 
The universal law  (\ref{current2})
for the boundary behavior of the current
holds for general BQFTs which are covariant, gauge invariant, unitary
and renormalizable, or equivalently, for BQFTs whose Weyl anomaly
is given by \eq{anom-4d}. 
In a material system, the $1/x$ dependence of the current is expected to be suppressed 
exponentially at large distances due to effects of dimensional operators. 
This can be verified with a holographic
calculation \cite{private}.  This has also been verified recently
by first principle calculation for the spatial component of the current \cite{Chernodub:2018ihb}. 
There is also an  interesting proposal to measure the beta function
by using the time component of the current \eq{current2} near the boundary of semimetals
\cite{Chernodub:2019blw}.

\subsection{Holographic boundary conformal field theory}

The above derivation is quantum field theoretic. The same result can also
be established from holography. BCFT
\cite{Cardy:2004hm,McAvity:1993ue} 
describes the fixed point of renormalization group flow in boundary
quantum field theory and has important
applications in  quantum field theory, string theory
and condensed matter system such as,
for example, renormalization
group flows and critical phenomena \cite{Cardy:2004hm} or
the topological
insulator \cite{Hasan:2010xy}. 
For general shape of the boundary, traditional perturbative analysis of BCFT
becomes exceedingly complicated. 
In addition to traditional
field theory techniques, see, e.g.
\cite{
Fursaev:2015wpa,Herzog:2015ioa,Miao:2017aba,Herzog:2017kkj,
Jensen:2017eof,
Kurkov:2017cdz,Kurkov:2018pjw,Rodriguez-Gomez:2017kxf},
the need of a
non-perturbative approach using symmetries 
or dualities is evident. 
A non-perturbative holographic dual description to BCFT was initiated by Takayanagi
in \cite{Takayanagi:2011zk} and later developed for general shape of boundary
geometry in \cite{Miao:2017gyt,Chu:2017aab}. The duality has been extensively
studied in the literature, with many interesting results obtained. See, for example
\cite{Fujita:2011fp,Nozaki:2012qd,Astaneh:2017ghi,
Flory:2017ftd,Bhowmick:2017egz,Herzog:2017xha,
Seminara:2017hhh,Chang:2018pnb,Seminara:2018pmr,Miao:2018qkc,Andrei:2018die}.

\begin{figure}[t]
\centering
\vspace{10pt}
\includegraphics[width=5cm]{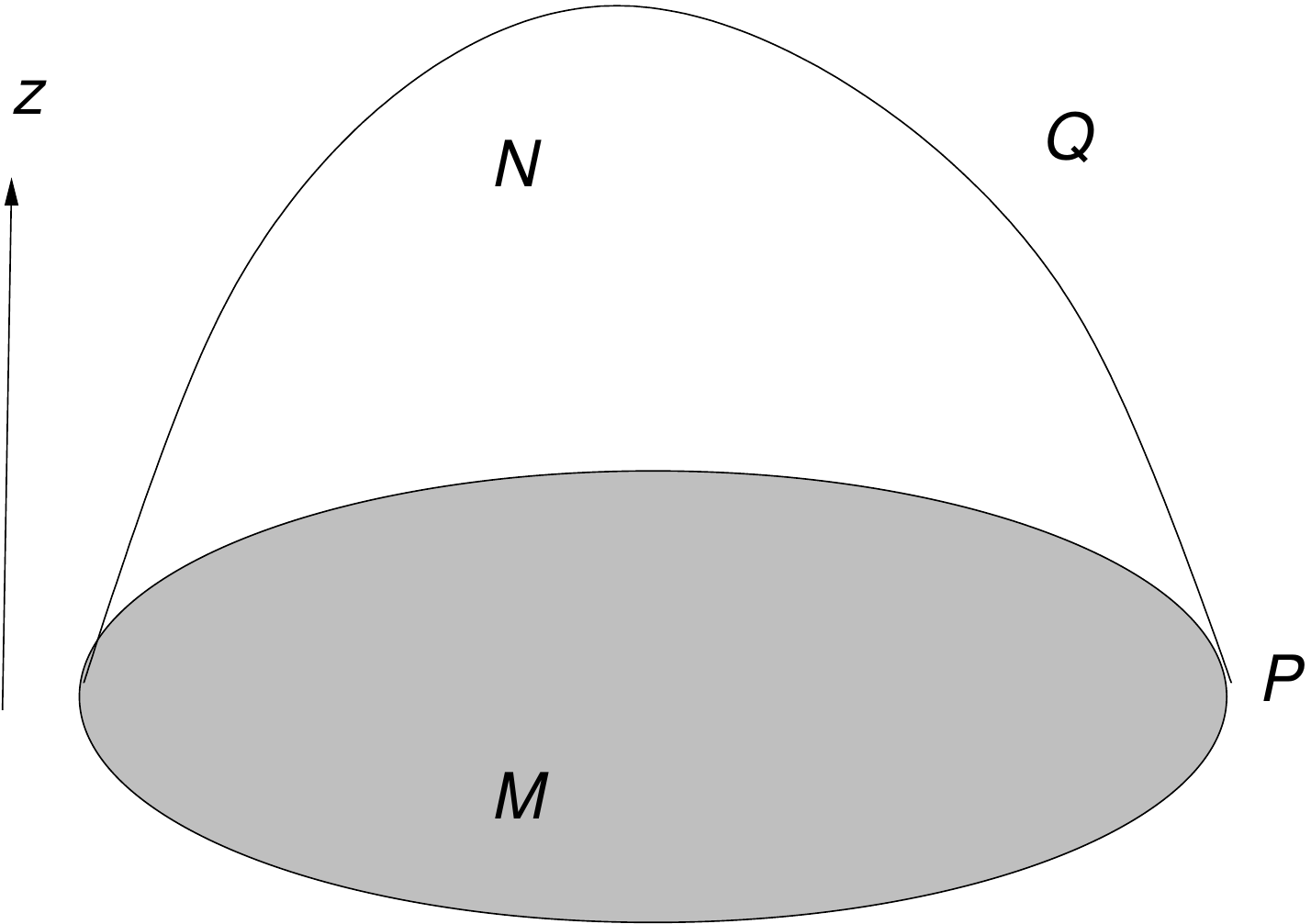}
\caption{BCFT on $M$ and its dual $N$}
\label{MNPQ}
\end{figure}

To investigate the renormalized current in holographic models
of BCFT,
let us add a $U(1)$ gauge field to the holographic model and
consider the following gauge invariant action for holographic BCFT 
 ($16\pi G_N =1$)
\begin{eqnarray}\label{action1}
  I&=&\int_N \sqrt{G} (R-2 \Lambda-\frac{1}{4}\mathcal{F}_{MN}^2
     ) +2\int_Q \sqrt{\gamma} (K-T).
\end{eqnarray}
Here $\mathcal{F}$ is the bulk field strength which reduces  to $F$
on the boundary $M$.
 The bulk indices are denoted by the capital Roman letters
 $L,M,N = 0, 1, \ldots, 4$
 and the indices of the four-dimensional manifolds $M$ and $Q$ are
 denoted by Greek letters $\mu, \nu$ etc. It should be clear from the context
 whether we are refering to the manifold $M$ or $Q$.
 As in standard AdS/CFT
 correspondence, the field $A_{\m}$ is completely
 arbitrary and does not need to
 satisfy any equation of motion.
 The constant
parameter $T$ is a measure of the boundary degree of freedom of the
BCFT. The holographic dual \eq{action} is defined once the shape of $N$ is
known. As
$N$ is of codimension one, the location of $Q$ is determined by a single
function. A 
  consistent model of holographic BCFT
  was found by  considering a mixed boundary conditions on $Q$ and the
  following  trace condition \cite{Miao:2017gyt,Chu:2017aab}
 \be
K = \frac{d}{d-1} T \label{NBC1-g}
\ee
was obtained.
The employment of a mixed boundary conditions  is a reasonable assumption
  if one think of $Q$ as a brane and then there should be a single embedding
  equation for it.
  In addition, we impose a Neumann boundary condition for the gauge field,
  \be
\mathcal{F}_{MN}\; n^M_Q\Pi^{N}_{\ \a}=0.
\label{NBC-A}
\ee
Here $n_Q$ is the inward-pointing normal vector on $Q$,
the beginning Greek letters $\a,\b$ etc denote indices on $Q$, and $\Pi$ is
the projection operator which gives the vector field and metric on $Q$:
$\bar{A}_{\a}=\Pi^{M}_{\ \a}\mathcal{A}_{M}$ and
$\gamma_{\alpha\beta}=\Pi^{M}_{\ \alpha}\Pi^{N}_{\ \beta}G_{MN}$.

  We note that \cite{Miao:2017aba} the manifold $N$ is actually singular
  since 
  the normal of $N$ is discontinuous at the junction $P$. Due to
  this discontinuity,
  an expansion in small $z$ in the form of
Fefferman--Graham (FG) asymptotic expansion \cite{AST_1985__S131__95_0}
  would not be  sufficient, and 
  one needs to have a full analytic control of the metric near $P$,
  i.e. near $z=0=x$. The need of a non-FG expanded bulk metric
was already anticipated in \cite{Nozaki:2012qd}.
The  general form of this non-FG expanded bulk metric
  that is analytic near $P$ 
  was  successfully constructed in \cite{Miao:2017aba} by
  considering
  an expansion in  small exterior curvature of the boundary surface $P$.
  Moreover it was found that \cite{Miao:2017aba}
 by using the non-FG
 expansion of the metric in the bulk, the tensor embedding equation
\be
K_{\alpha\beta}-(K-T)\gamma_{\alpha\beta} =0 \label{NBC-g}
\ee
for $Q$ as proposed originally by Takayanagai \cite{Takayanagi:2011zk}
is also  consistent: with the tensor model \eq{NBC-g}
considered as a special case of the scalar model \eq{NBC1-g}.
    
Now back to our system. Let us denote the five-dimensional
bulk indices by
$S =(z,\mu)$, and the six-dimensional field theory indices by
$\mu =(x,a)$ with
$a= 0,1, \ldots, 2$.
For simplicity, let us consider the case of a  flat half space $x\geq 0$.
The bulk metric reads
\begin{equation}\label{AdSmetric}
{\rm d}s^2=R^2 \frac{{\rm d}z^2+{\rm d}x^2+\delta_{ab}{\rm d}y^a{\rm d}y^b}{z^2}.
\end{equation}
In this case, \eq{NBC-g} reduces to \eq{NBC1-g}, and $Q$ is given by
\cite{Takayanagi:2011zk}
\begin{equation}\label{Q}
\ x=-z \sinh (\rho/R),
\end{equation}
where we have reparametrized $T=3 \tanh \rho$. As for
the solution for the vector field,  due to the
planar symmetry of the boundary, we consider $A_{\mu}$ that depends
only on the coordinates $z$ and $x$. The Maxwell equations
$\nabla_{\mu}\mathcal{F}^{\mu\nu}=0$ can be solved with
$\mathcal{A}_z=\mathcal{A}_z(z)$, $\mathcal{A}_x=\mathcal{A}_x(x)$ and
$\cA_a$ satisfying,
\begin{eqnarray}\label{EOMvector}
z \del_x^2 \mathcal{A}_a -
\del_z \mathcal{A}_a+z \del_z^2 \mathcal{A}_a=0.
\end{eqnarray}
One can solve the above equation by separation of
variables $\mathcal{A}_a(z,x)=Z(z)X(x)$, and then substitute the general solutions
to (\ref{NBC-A}) to obtain the solution by brute force.  However there is a
quicker trick.
Inspired by similar considerations in \cite{Miao:2017aba},  
let us take the following ansatz for the vector field
\begin{eqnarray}\label{vectoransatz}
  \mathcal{A}_a= \sum_{n=0}  x^n f_n\left(\frac{z}{x}\right) A^{(n)}_a,
\end{eqnarray}
where we set $f_i (0)=1$ so that $\mathcal{A}_a$ reduce
to the gauge field $A_a$  at the AdS boundary $z=0$. Here 
$A^{(i)}$ are the expansion coefficients of $A_a$ about the boundary: 
\begin{eqnarray}\label{vectorgauge}
A_a=\sum_{m=0} x^n A^{(n)}_a .
\end{eqnarray}
In particular, $A_a^{(1)}$ is given by the field strength at the boundary:
\be
\label{AF}
A_a^{(1)} = F_{xa} = F_{na}.
\ee
Note that
in the derivative expansions we have $O(A^{(i)}) \sim O(\partial)^i$.
Substituting (\ref{vectoransatz}) into (\ref{EOMvector}) we get 
\begin{eqnarray}
s (s^2+1) f_1''(s)-f_1'(s)=0, 
\end{eqnarray}
at the linear order $O(\partial)$. Recall that $f_1(0)=1$,
we have the solution $f_1(s)=1-c_1+c_1 \sqrt{1+s^2}$, and
(\ref{vectoransatz}) reads
\begin{eqnarray}\label{vectoransatz1}
\mathcal{A}_a=A^{(0)}_a+  \left( (1-c_1) x+ c_1\sqrt{x^2+z^2} \right) A^{(1)}_a,
\end{eqnarray}
where we have ignored the higher order terms since they are irrelevant
to the current (\ref{current2}) of order $O(\partial)$, or equivalently,  $O(F)$.
Note also that we have analytic continuated $ x \sqrt{1+\frac{z^2}{x^2}}$ to
$ \sqrt{x^2+z^2}$ in order to get smooth solution at $x=0$. 
Imposing the boundary condition (\ref{NBC-A}) on $Q$, we get 
$c_1=1$. One can check directly that the solution $\mathcal{A}_z=\mathcal{A}_z(z),
\mathcal{A}_x=\mathcal{A}_x(x)$ and 
\begin{eqnarray}\label{vectors}
\mathcal{A}_a=A^{(0)}_a+ A^{(1)}_a \sqrt{x^2+z^2}
\end{eqnarray} 
is  indeed an exact solution to the
Maxwell equations and the boundary condition (\ref{NBC-A}) in AdS. 

From the gravitational action (\ref{action1}), we can derive the holographic
current  \cite{Chu:2018ksb}
\begin{eqnarray}\label{holocurrent0}
  \la J^a \ra=\lim_{z\to 0}\frac{\delta I}{\delta A_a}
  =\lim_{z\to 0}\sqrt{G}\mathcal{F}^{za}
\end{eqnarray}
Substituting the solutions (\ref{AdSmetric}), (\ref{vectors})
into (\ref{holocurrent0}), we obtain
\begin{eqnarray}\label{holocurrent}
\la J_a \ra=\partial_z^2\mathcal{A}_a|_{z=0}= -\frac{F_{an} }{x} +O(1),
\end{eqnarray}
where we have used \eq{AF}. 
On the other hand, the holographic Weyl anomaly of (\ref{action1}) is
obtained in \cite{Genolini:2016ecx} with the central charge given by
\begin{eqnarray}\label{holocharge}
b_1=-\frac{R^3}{16 \pi G_N}.
\end{eqnarray}
Now it is clear that the holographic BCFT satisfies the universal law
of current (\ref{current2}).  
It is remarkable that current (\ref{holocurrent}) is independent of the
parameter $T$ , which shows that
near-boundary current for 4d BCFT is indeed independent of boundary conditions.

\subsection{Induced string current}

Let us consider a six-dimensional BCFT with gauge symmetry
defined on a manifold $M$.
The Yang--Mills gauge
field is not conformal invariant in six-dimensions,
instead a 2-form gauge field $B_{\m\n}$ is. 
For simplicity, we consider Abelian gauge field here. 
The 2-form gauge potential is naturally coupled to the world-sheet
$\S$ of a string
with the  minimal coupling
\be \label{IB}
I_B = \int_\S B = \int_M J^{\m\n} B_{\m\n}  
\ee
where 
\be
J^{\m\n} = \l \e^{\a\b} \frac{\del X^\m}{\del \s^\a} \frac{\del X^\n}{\del \s^\b}
\d^{(4)} (X - X(\s^a))
\ee
is a two-form string current that arises from the
motion of the string and
$\l$ is the string charge density.
Next let us introduce a boundary $P = \del M$. This breaks the bulk conformal
symmetry and the one point function of the current can become nontrivial now. 
As the current
$J_{\m\n}$ has a mass dimension 4,  the vacuum expectation value of
the renormalized current generally takes the form
\be \label{J-asym}
\la J_{\m\n} \ra
= \frac{1}{x} J_{\m\n}^{(1)} + \log x J_{\m\n}^{(0)} + \cdots 
\ee
near the boundary.
Here we have used  gauge invariance and the conservation law 
\be \label{DJ}
D_\m J^{\m\n} =0
\ee
to rule out  terms like
$ J_{\m\n}^{(4)}/x^4,  J_{\m\n}^{(3)}/x^3,  J_{\m\n}^{(2)}/x^2$. In \eq{J-asym}, 
$\cdots$ denotes terms that are regular at $x=0$, and 
$J_{\m\n}^{(1)}$ and $J_{\m\n}^{(0)}$ are functions of
dimension 3 and 4 respectively.
Their form are constrained by
\eq{DJ} and the Lorentz and gauge symmetries of the theory.
For example, one
can easily determine that
\be \label{J1}
\begin{aligned}
J_{\m\n}^{(1)} &= \a_1 H_{\m\n \l} n^\l + \a_2 \cD_{[ \m} \cD_{\n ]} k\,+\\
&\kern.5cm+\, \a_3 \cD_{[ \m} \cD_\l k^\l{}_{\n ]} +\a_4 \cD_\l \cD_{[ \m} k^\l{}_{\n ]},
\end{aligned}
\ee
where $H_{\m\n \l}, n_\m, \cD_\m, k_{\m\n}$ are respectively the background
3-form field
strength, normal vector to the boundary, induced covariant derivative and the
extrinsic curvature of the boundary.
The coefficients $\a_i$ are arbitrary and contain important physical
information of the theory. Unlike the four dimensional
case, the
background gauge field part of the Weyl anomaly is unknown. Therefore, instead of
trying to determine the induced current in terms of the Weyl anomaly, 
let us
proceed first with the holographic analysis and determine the
near boundary current using boundary holography.

For six-dimensional BCFT, we include a
2-form potential $B$ in the bulk whose boundary value
is dual to the string current $J_{\m\n}$,
the holographic dual of the BCFT is described by the gravitational action
\be \label{action}
I =
\frac{1}{16 \pi G_N}
\int {\rm d}^7 x \sqrt{-G} \left(R-2 \Lambda - \frac{1}{6} \cH_{LMN}^2\right).
  \ee
  Here $G_N$ is the Newton constant in 7 dimensions
 and $\cH = d \cB$. $\cB$ is the bulk gauge field whose boundary value is
 given by the gauge field $B$ on the boundary $M$. The induced
 current can be calculated in a similar way as above and we obtain \cite{Chu:2018fpx}
\be \label{J-B}
\la J^{ab} \ra = \lim_{z\to 0} \frac{\d I }{\d \cB_{ab}}
= b  \frac{ H_{xab}}{x},
\ee
where
$b = - \frac{R^5}{16 \pi G_N}$ is a constant. As in the four dimensional case, the 
current \eq{J-B} is independent of boundary condition.

\section{Weyl anomaly and induced current for six-dimensional CFT}

In the above, we have shown that for a four-dimensional BCFT with a $U(1)$
gauge symmetry generated by a gauge field $A_\m$, the 
relation \eq{key} implies the existence of  an induced current, and determine its form near boundary in terms of the Weyl anomaly of the theory.
In general the Weyl anomaly can be computed from the quantum effects
of matter loops on the path integral with external gauge fields.

In higher dimensions, the gauge field contribution to the Weyl anomaly is
unknown.
Nevertheless,
even without any knowledge of the path integral or how the higher rank gauge
field,
one can establish
a similar relation \eq{key} between the Weyl anomaly and the boundary current.
couples to the other fields of the system. For example in $d=6$, we have
the relation \cite{Chu:2018fpx}
\be \label{key1}
(\d \mathcal{A} )_{\del M_\e} = \Big(\int_{M_\e} \sqrt{g} J^{\m\n} \d B_{\m\n}
\Big)_{\log \frac{1}{\e}}.
\ee
Using the holographic result \eq{J-B},
one can verify that \eq{key} is satisfied
with $\mathcal{A}$ given by
\be \label{anom-6d}
\mathcal{A}=\int_M \sqrt{g}\; \frac{b}{6} H_{\m\n\l}^2
\ee
Note that one can also use the AdS/BCFT to compute the holographic stress tensor
and the Weyl anomaly \cite{Miao:2017gyt,Chu:2017aab}.
The same result is obtained.
This is our prediction for the form
of the gauge field contribution in the Weyl anomaly in 6d CFT with tensor gauge
field.

A particularly interesting setting where a tensor gauge potential $B_{\m\n}$ appears
is in the theory of multiple M5-branes.
The maximal supersymmetric M5-brane has (2,0) supersymmetry and admits a
self dual tensor multiplet with a self-dual gauge potential $B_{\m\n}$.
For this theory, we can work out the
value of $b$ in the current \eq{anom-6d}. 
The holographic dual of a system of $N$ coincident M5-branes is given
by M-theory on the AdS$_7 \times S^4$ background with a constant 4-form field
strength and the metric
\be
{\rm d}s^2 = R^2 \frac{{\rm d}z^2 + {\rm d}x_6^2}{z^2} + R'{}^2 {\rm d}\Omega_4^2
\ee
where $R' =R/2$, $R = 2 (\pi N)^{1/3} l_{11}$ and
$l_{11}$ is the 11-dimensional Planck
length. Since the 7-dimensional Newton  $G_N = G_N^{(11)}/{\rm Vol} (S^4)$ and
$G_N^{(11)} = 16 \pi^7 l_{11}^9$, we obtain
\be \label{b-6d}
b = - \frac{N^3}{3 \pi^3}.
\ee
The tensor multiplet obeys a nonlinear self-duality relation
\cite{Howe:1997vn} and the
anomaly \eq{anom-6d} is non-trivial.
We note that in four-dimensions, the coefficient $b$ is given by the beta function of
the theory and
is proportional to the number of degree of freedom that couple to the
$U(1)$ gauge field. 
Here we expect that $b$ to be proportional to the degrees of freedom that
couple to the 2-form gauge field. Our result \eq{b-6d} suggests that
the number of degree of freedom in the (2,0) theory is proportional to
$N^3$ for large $N$. We note that a factor of $N^3$ also appear in the
entropy of a system of coincident near extremal black 5-branes solution
\cite{Klebanov:1996un}.
However we emphasis that the associated physical mechanism is different:
here there is no horizon in the geometry and a different observable,
a conserved current, is considered.

\section{Open problems}

In view of the research presented here, there are a number of questions that one can ask naturally.

Is it possible and how to derive the result of Weyl anomaly from a fundamental quantum 
computation? D-branes in the presence of a constant 2-form NS-NS $B$-field background is 
described by a non-commutative geometry of Moyal type. This can be derived by considering open string quantisation. For a M5-brane in the presence of a constant 3-form $C$-field 
background, there must be some kind of non-commutative geometry. However it is completely 
mysterious. All we know is that it must reduce to a Moyal type non-commutative geometry
upon a dimensional  reduction. It may be possible to learn more about this geometry by establishing a link between the quantum geometry on the brane with the magnetic Casimir current, both  consequence of the background flux.

\bibliography{allbibtex}

\bibliographystyle{prop2015}

\end{document}